\documentclass[aps,PRB,twocolumn,superscriptaddress,preprintnumbers]{revtex4-2}
\usepackage[T1]{fontenc}
\usepackage{times}
\usepackage{graphicx,color}
\usepackage{amsfonts,amsmath,amssymb,amsbsy}
\usepackage[colorlinks=true, linkcolor=blue, citecolor=blue, urlcolor=blue]{hyperref}
\usepackage{float}

\begin{document}

\title{Out-of-plane Edelstein effects: Electric-field induced magnetization
in $p$-wave magnets}
\author{Motohiko Ezawa}
\affiliation{Department of Applied Physics, The University of Tokyo, 7-3-1 Hongo, Tokyo
113-8656, Japan}

\begin{abstract}
In-plane magnetization is induced by the Edelstein effect in the Rashba
spin-orbit interaction system. However, out-of-plane magnetization is more
useful for switching a ferromagnetic memory. We study analytically and
numerically electric-field induced magnetization in $p$-wave magnets with
the aid of the Rashba interaction based on a simple two-band model. The
out-of-plane magnetization is induced when the N\'{e}el vector of the $p$%
-wave magnet is along the $z$ direction. We also show that no magnetization
is induced in the absence of the Rashba interaction. The electric-field
induced magnetization will be useful for future switching technology of
ferromagnetic memories based on the $p$-wave magnet.
\end{abstract}

\date{\today }
\maketitle

\begin{figure}[t]
\centerline{\includegraphics[width=0.49\textwidth]{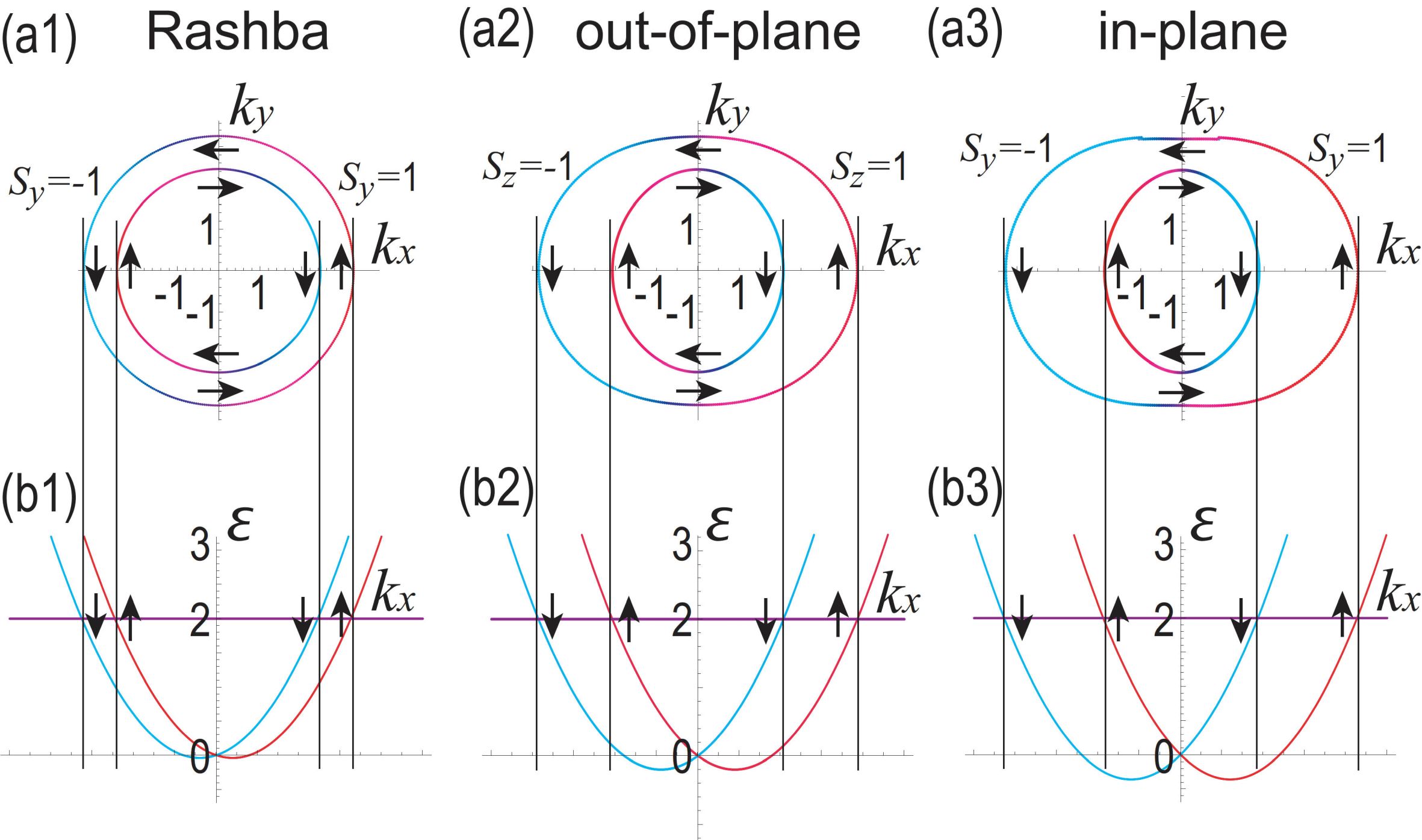}}
\caption{(a1), (a2) and (a3) Fermi surfaces at $\protect\mu =2\protect%
\varepsilon _{0}$ in the $(k_{x},k_{y})$ plane. The momentum axis is in
units of $k_{0}$. (b1), (b2) and (b3) Energy spectrum along the $x$ axis.
The vertical axis is the energy $\protect\varepsilon $ in units of $\protect%
\varepsilon _{0}$, while the horizontal axis is $k_{x}$ in units of $k_{0}$.
Purple lines represent $\protect\mu =2\protect\varepsilon _{0}$. (a1) and
(b1) Rashba model. (a2) and (b2) Out-of-plane $p$-wave magnet. (a3) and (b3)
In-plane $p$-wave magnet. Red color represents $S_{y}=1$, while cyan color
represents $S_{y}=-1$ in (a1), (b1), (a3) and (b3). On the other hand, red
color represents $S_{z}=1$, while cyan color represents $S_{z}=-1$ in (a2)
and (b2). Arrows represent the spin direction. We have set $m=4\hbar
^{2}k_{0}^{2}/\protect\varepsilon _{0}$, $J=0.4\protect\varepsilon %
_{0}/k_{0} $ and $\protect\lambda =0.2\protect\varepsilon _{0}/k_{0}$. }
\label{FigSurface}
\end{figure}

\textbf{Introduction:} Multiferroic effects are cross correlation effects in
electromagnetic responses\cite{Fiebig,Eeren,Ramesh}. One is the electric
polarization induced by magnetic-field\cite{Kimura,Hur,Nan} and the other is
the magnetization induced by electric field\cite{Ohno,Chu,Heron,Matsukura}.
Especially, the latter is important for the flip of a magnetic memory under
the control of the external electric field. The Edelstein effect\cite%
{Edel,Gamba} is a promising way for it. It is based on the spin-momentum
locked Fermi surface induced by the Rashba interaction\cite{Manchon}. In the
presence of the Rashba interaction, the spin direction depends on the angle
of the momentum as shown in Fig.\ref{FigSurface}(a), where the Fermi surface
forms a perfect circle. By applying in-plane electric field, the Fermi
distribution is shifted, which induces the imbalance in the spin
accumulation. It results in the net in-plane magnetization. However,
out-of-plane magnetization is desirable for practical applications to
ferromagnetic memories because ferromagnet points to out-of-plane direction.
The Edelstein effect is confirmed in various experiments\cite%
{Sanchez,Kawasuso,Du}. It is also realized in semiconductors\cite%
{Kato,Peters}, Rashba-Dresselhaus systems\cite{Bryk,Monte}, topological
insulators\cite{Mell,SanchezPRL}, Weyl semimetals\cite{Johan}.

Recently, a $p$-wave magnet was proposed based on first-principle
calculations\cite{pwave}, where the Fermi surface has the $p$-wave symmetry.
A material candidate is CeNiAsO. The system is described by a four-band
tight-binding model. Subsequently, the Edelstein effect without the Rashba
interaction was proposed in this magnet\cite{Chak}.

In this paper, we investigate the Edelstein effect in a two-dimensional
system made of a $p$-wave magnet with the Rashba interaction, which is
described by a simple two-band continuum model. By applying an in-plane
electric field, we demonstrate the emergence of the out-of-plane
magnetization provided the N\'{e}el vector of the $p$-wave magnet is taken
perpendicular to the plane. Namely, it is possible to switch the
out-of-plane magnetization by controlling the external in-plane electric
field. It would be useful for the switch of a ferromagnetic memory. Our
results will open a new way to the magnetic-memory switching device by
electric field.

\textbf{Edelstein effect:} The magnetization is induced by the change of the
Fermi distribution in the Rashba system, which is called the Edelstein
effect. We study such an effect by controlling external electric field $%
\mathbf{E}$. We expand the Fermi distribution in powers of $\mathbf{E}$, $%
f=f^{\left( 0\right) }+f^{\left( 1\right) }+\cdots $, where$f^{\left(
0\right) }=1/\left( \exp \left[ \left( \varepsilon -\mu \right) /k_{B}T%
\right] +1\right) $ is the Fermi distribution function, $\mu $ is the
chemical potential, $\varepsilon $ is the energy, and $f^{\left( 1\right) }$
is the first-order perturbated Fermi distribution function as a function of $%
\mathbf{E}$. By solving the Boltzmann equation, it is explicitly given by 
\begin{equation}
f^{\left( 1\right) }=\frac{e\tau }{\hbar }\mathbf{E\cdot }\nabla _{\mathbf{k}%
}f^{\left( 0\right) },
\end{equation}%
where $\tau $ is the relaxation time. We focus on the linear response.

The magnetization is given by the formula\cite%
{Garete,Garete2,Zele,Chak,Li2015,ManchonRMP,Tenzin,Her,Chen,Trama,Hu,Pari}%
\begin{equation}
\mathbf{M}=\frac{e\tau }{\hbar }g\mu _{\text{B}}\int \frac{d^{3}k}{\left(
2\pi \right) ^{3}}\mathbf{S}\left( \mathbf{k}\right) f^{\left( 1\right) },
\end{equation}%
where $\mathbf{S}\left( \mathbf{k}\right) \equiv \left\langle \psi _{\mathbf{%
k}}\right\vert \mathbf{\sigma }\left\vert \psi _{\mathbf{k}}\right\rangle $
is the expectation value of the spin operator $\mathbf{\sigma }$, $g$ is the
g factor, and $\mu _{\text{B}}$ is the Bohr magneton. The magnetization
formula is rewritten as%
\begin{equation}
\mathbf{M}=\frac{e\tau }{\hbar }g\mu _{\text{B}}\int \frac{d^{3}k}{\left(
2\pi \right) ^{3}}\mathbf{S}\left( \mathbf{k}\right) \left( \mathbf{E\cdot }%
\nabla _{\mathbf{k}}\varepsilon \right) \frac{\partial }{\partial
\varepsilon }f^{\left( 0\right) }.
\end{equation}%
At zero temperature, it is given by%
\begin{equation}
\mathbf{M}=e\tau g\mu _{\text{B}}\int \frac{d^{3}k}{\left( 2\pi \right) ^{3}}%
\mathbf{S}\left( \mathbf{k}\right) \left( \mathbf{E\cdot v}\right) \delta
\left( \varepsilon _{\mathbf{k}}-\mu \right) ,
\end{equation}%
where we have defined the velocity operator $\hbar \mathbf{v}=\nabla _{%
\mathbf{k}}\varepsilon $.

We apply the electric field along the $x$ axis in the two-dimensional
system. When there are two Fermi surfaces $k_{\pm }\left( \phi \right) $\
around the origin as illustrated in Fig.\ref{FigSurface}, it is rewritten as 
\begin{equation}
\mathbf{M}=\frac{e\tau g\mu _{\text{B}}E_{x}}{\left( 2\pi \right) ^{2}W}%
\sum_{\pm }\int kdkd\phi \mathbf{S}\left( \mathbf{k}\right) v_{x}\frac{%
\delta \left( k_{\pm }\left( \phi \right) -k\right) }{\left\vert \frac{%
\partial \varepsilon }{\partial k}\right\vert },  \label{Edel}
\end{equation}%
or%
\begin{equation}
M_{i}=e\tau \alpha _{ij}^{\text{ME}}E_{j},
\end{equation}%
in terms of the magnetoelectric susceptibility $\alpha _{ij}^{\text{ME}}$
given by%
\begin{equation}
\alpha _{ix}^{\text{ME}}=\frac{g\mu _{\text{B}}}{\left( 2\pi \right) ^{2}W}%
\sum_{\pm }\int d\phi \left. \frac{kv_{x}S_{i}\left( \mathbf{k}\right) }{%
\left\vert \frac{\partial \varepsilon }{\partial k}\right\vert }\right\vert
_{k=k_{\pm }\left( \phi \right) },  \label{ME}
\end{equation}%
where $k_{\pm }\left( \phi \right) $ is obtained by solving $\varepsilon
\left( k_{\pm }\left( \phi \right) \right) =\mu $, $v_{x}=\frac{\partial
\varepsilon }{\hbar \partial k_{x}}$ is the velocity along the $x$ axis, $%
W^{-1}\equiv \int dk/\left( 2\pi \right) $ is the width of the sample, and
we have introduced the polar coordinate $k_{x}=k\cos \phi $, $k_{y}=k\sin
\phi $.

\begin{figure}[t]
\centerline{\includegraphics[width=0.44\textwidth]{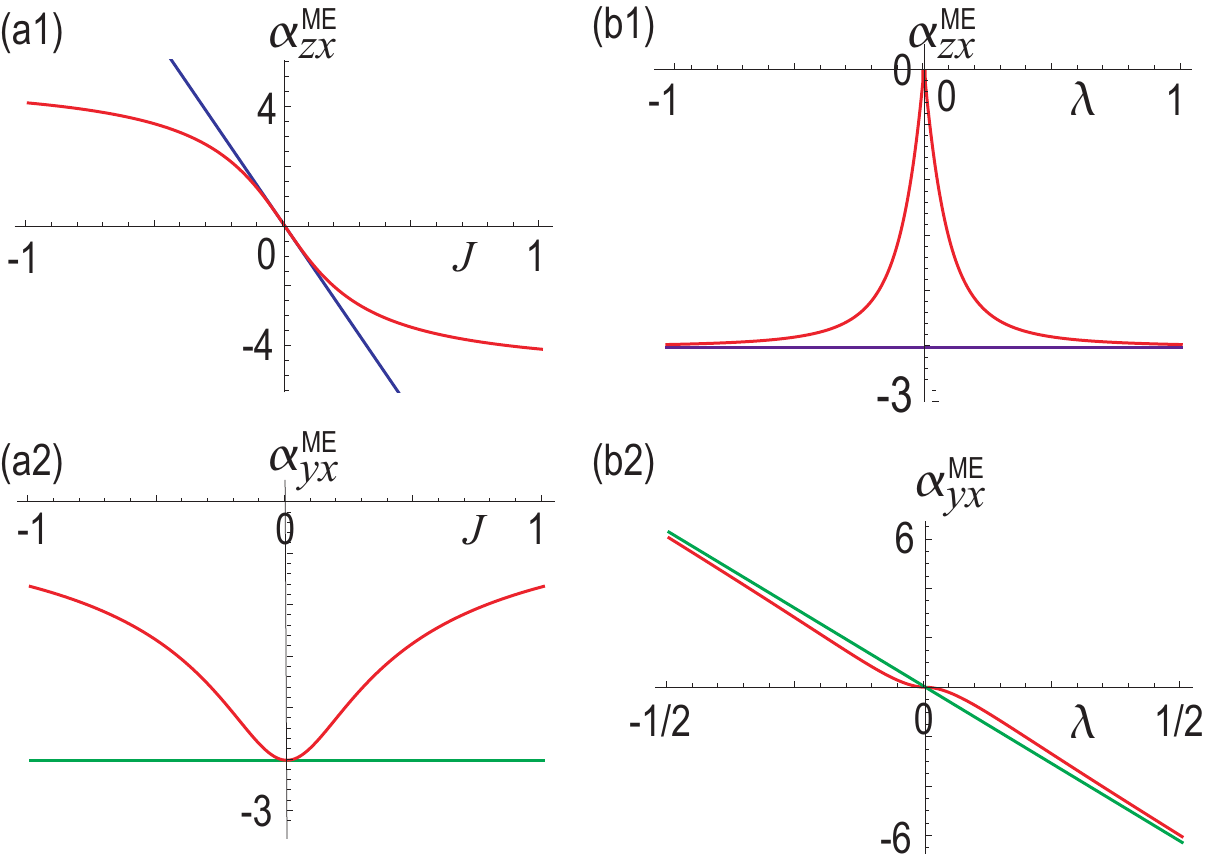}}
\caption{Out-of-plane Edelstein effect. (a1) $\protect\alpha _{zx}^{\text{ME}%
}$ and (a2) $\protect\alpha _{yx}^{\text{ME}}$ as a function of $J$. We have
set $\protect\lambda =0.2\protect\varepsilon _{0}/k_{0}$. (b1) $\protect%
\alpha _{zx}^{\text{ME}}$ and (b2) $\protect\alpha _{yx}^{\text{ME}}$ as a
function of $\protect\lambda $. We have set $J=0.2\protect\varepsilon %
_{0}/k_{0}$. Red curves represent the numerical results without using the
perturbation theory. Blue and green lines represent the analytic results for 
$\protect\alpha _{zx}^{\text{ME}}$ and $\protect\alpha _{yx}^{\text{ME}}$ in
Eq.(\protect\ref{SzL}), respectively. $\protect\alpha _{xx}^{\text{ME}}$, $%
\protect\alpha _{yx}^{\text{ME}}$ and $\protect\alpha _{zx}^{\text{ME}}$ are
in units of $\frac{g\protect\mu _{\text{B}}k_{0}}{\hbar W}$, while $J$ and $%
\protect\lambda $ are in units of $\protect\varepsilon _{0}/k_{0}$. We have
set $\protect\mu =2\protect\varepsilon _{0}$ and $m=4\hbar ^{2}k_{0}^{2}/%
\protect\varepsilon _{0}$. }
\label{FigEdelZ}
\end{figure}

\textbf{Out-of-plane }$p$\textbf{-wave Edelstein effects:} We consider a
two-dimensional system made of the $p$-wave magnet with the Rashba
interaction. The simplest model is a two-band model consisting of the
free-electron term, the Rashba term\cite%
{SmejRev,SmejX2,Zu2023,Sun,Diniz,Rao,Amund,EzawaAlter,EzawaPNeel} and the $p$%
-wave term\cite{pwave,Maeda,EzawaPwave,Brek,EzawaPNeel,GIAlter}, which is
realized as a heterostructure of a magnet on a substrate. The Hamiltonian is
given by\cite{EzawaPNeel},%
\begin{equation}
H\left( \mathbf{k}\right) =\frac{\hbar ^{2}\left( k_{x}^{2}+k_{y}^{2}\right) 
}{2m}\sigma _{0}+\lambda \left( k_{x}\sigma _{y}-k_{y}\sigma _{x}\right)
+Jk_{x}\mathbf{n}\cdot \mathbf{\sigma },  \label{pHamil}
\end{equation}%
where $m$ is the effective mass of free electrons, $\lambda $ is the
magnitude of the Rashba interaction, and $\mathbf{n}$ is the $p$-wave N\'{e}%
el vector with $J$ its magnitude. The Rashba interaction is introduced by
placing a sample on the substrate. We study the case where the N\'{e}el
vector is along the $z$ direction, $\mathbf{n=}\left( 0,0,1\right) $.

\begin{figure}[t]
\centerline{\includegraphics[width=0.48\textwidth]{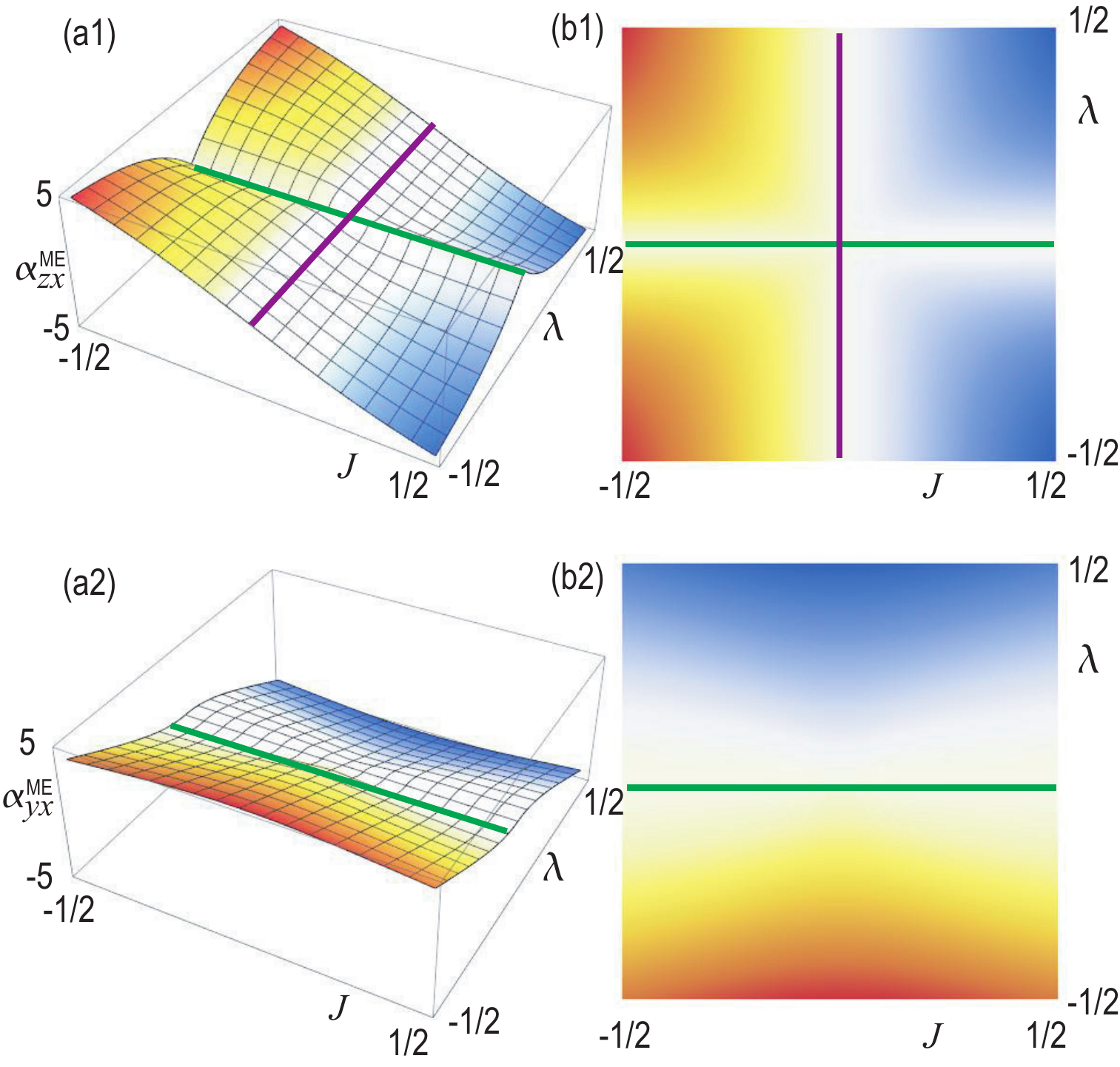}}
\caption{Out-of-plane Edelstein effect. (a1) Bird's eye's view of $\protect%
\alpha _{zx}^{\text{ME}}$ in the $J$-$\protect\lambda $ plane, and (b1) its
color plot. (a2) Bird's eye's view of $\protect\alpha _{yx}^{\text{ME}}$ in
the $J$-$\protect\lambda $ plane, and (b2) its color plot. Green line
represents the condition $\protect\lambda =0$ and the purple line represents
the condition $J=0$. See the caption of Fig.2 for the units of various
variables. We have set $\protect\mu =2\protect\varepsilon _{0}$ and $%
m=4\hbar ^{2}k_{0}^{2}/\protect\varepsilon _{0}$. }
\label{FigPhaseZ}
\end{figure}

The energy spectrum is given by%
\begin{equation}
\varepsilon _{\pm }=\frac{\hbar ^{2}k^{2}}{2m}\pm k\sqrt{F\left( \phi
\right) },
\end{equation}%
with%
\begin{equation}
F\left( \phi \right) \equiv \lambda ^{2}+J^{2}\cos ^{2}\phi .
\end{equation}%
The expectation value of the spin operator $\mathbf{S}\left( \mathbf{k}%
\right) $ is determined as 
\begin{equation}
S_{x}^{\pm }=\mp \frac{\lambda \sin \phi }{\sqrt{F\left( \phi \right) }}%
,\quad S_{y}^{\pm }=\pm \frac{\lambda \cos \phi }{\sqrt{F\left( \phi \right) 
}},\quad S_{z}^{\pm }=\pm \frac{J\cos \phi }{\sqrt{F\left( \phi \right) }},
\end{equation}%
which does not depend on $k$ and shown in Fig.\ref{FigSurface}(a3).

The integrand of Eq.(\ref{Edel}) is given analytically. By integrating it
over $\phi $ numerically, we obtain the susceptibility $\alpha _{ix}^{\text{%
ME}}$. We show the components $\alpha _{zx}^{\text{ME}}$ and $\alpha _{yx}^{%
\text{ME}}$ in Fig.\ref{FigEdelZ}, while $\alpha _{xx}^{\text{ME}}=0$. 

We derive an analytic formula for the susceptibility up to the first-order
in $J$ and $\lambda $. There are two Fermi surfaces $k_{\pm }\left( \phi
\right) $ as shown in Fig.\ref{FigSurface}(a). The outer Fermi surface $%
k_{-}\left( \phi \right) $ and the inner Fermi surface $k_{+}\left( \phi
\right) $ are given by%
\begin{equation}
\hbar k_{\pm }\left( \phi \right) =\sqrt{2\mu m}\mp mJ\cos \phi ,
\end{equation}%
which are shown in color in Fig.\ref{FigSurface}(a). The $y$ spin
polarization is maximized along the $x$ axis ($\phi =0,\pi $). 

The susceptibility (\ref{ME}) is calculated as%
\begin{equation}
\alpha _{xx}^{\text{ME}}=0,\quad \alpha _{yx}^{\text{ME}}=-\frac{g\mu _{%
\text{B}}m}{2\pi \hbar ^{3}W}\lambda ,\quad \alpha _{zx}^{\text{ME}}=-\frac{%
g\mu _{\text{B}}m}{2\pi \hbar ^{3}W}J  \label{SzL}
\end{equation}%
for $\mu >0$\ up to the first order in $J$\ and $\lambda $, which are shown
as the blue and green lines in Fig.\ref{FigEdelZ}, respectively.

We show the numerically obtained susceptibility in the $J$-$\lambda $ plane
in Fig.\ref{FigPhaseZ}. Eq.(\ref{SzL}) agrees well with the numerical result
for small $J/\lambda $. We find that $\alpha _{zx}^{\text{ME}}=0$\ along the
lines $J=0$\ and $\lambda =0$ as in Fig.\ref{FigPhaseZ}(b1), while $\alpha
_{yx}^{\text{ME}}=0$ along the line $\lambda =0$. The susceptibility $\alpha
_{zx}^{\text{ME}}$\ is positive for $J<0$\ and negative for $J>0$, where the
sign of $\alpha _{zx}^{\text{ME}}$\ does not depend on $\lambda $. On the
other hand, $\alpha _{yx}^{\text{ME}}$\ is positive for $\lambda <0$\ and
negative for $\lambda >0$, where the sign of $\alpha _{yx}^{\text{ME}}$\
does not depend on $J$.

\begin{figure}[t]
\centerline{\includegraphics[width=0.44\textwidth]{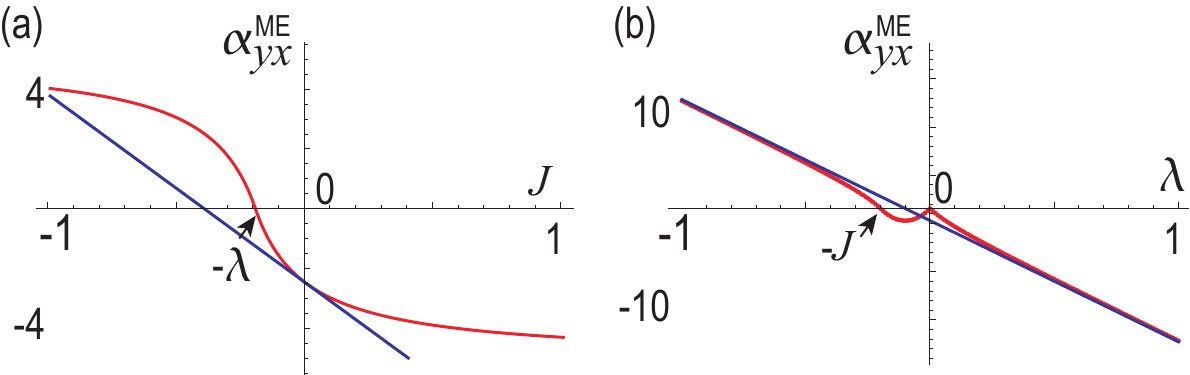}}
\caption{In-plane Edelstein effect. (a) $\protect\alpha _{yx}^{\text{ME}}$
as a function of $J$. We have set $\protect\lambda =0.2\protect\varepsilon %
_{0}/k_{0}$. b) $\protect\alpha _{yx}^{\text{ME}}$ as a function of $\protect%
\lambda $. We have set $J=0.2\protect\varepsilon _{0}/k_{0}$. Red curves
represent the numerical results without using the perturbation theory. Blue
lines represent the analytic results Eq. (\protect\ref{S1})  based on the
perturbation theory. See the caption of Fig.2 for the units of various
variables. We have set $\protect\mu =2\protect\varepsilon _{0}$ and $%
m=4\hbar ^{2}k_{0}^{2}/\protect\varepsilon _{0}$.}
\label{FigEdel}
\end{figure}

\textbf{In-plane }$p$\textbf{-wave magnet:} We next study the case where the
N\'{e}el vector is along the $y$ direction. The energy spectrum is given by%
\begin{equation}
\varepsilon _{\pm }=\frac{\hbar ^{2}k^{2}}{2m}\pm k\sqrt{G\left( \phi
\right) },
\end{equation}%
with%
\begin{equation}
G\left( \phi \right) \equiv \lambda ^{2}\sin ^{2}\phi +\left( J+\lambda
\right) ^{2}\cos ^{2}\phi .
\end{equation}

The Fermi surface is shown in Fig.\ref{FigSurface}(a3). There are two Fermi
surfaces $k_{\pm }\left( \phi \right) $. The outer Fermi surface $%
k_{-}\left( \phi \right) $ and the inner Fermi surface $k_{+}\left( \phi
\right) $ are analytically given by%
\begin{equation}
\hbar k_{\pm }\left( \phi \right) =\sqrt{2\mu m}\mp m\lambda +J\left( m\mp 
\frac{m\sqrt{m}\lambda }{\sqrt{2\mu }}\right) \cos ^{2}\phi
\end{equation}%
up to the first order in $J$ and $\lambda $. The spin direction is 
\begin{equation}
S_{\pm }^{y}=\pm \frac{\left( J+\lambda \right) x}{\sqrt{G\left( \phi
\right) }}.
\end{equation}

The susceptibility (\ref{ME}) is calculated as%
\begin{equation}
\alpha _{yx}^{\text{ME}}=-\frac{mg\mu _{\text{B}}}{4\pi \hbar ^{3}W}\left(
J+2\lambda \right) ,\qquad \alpha _{xx}^{\text{ME}}=\alpha _{zx}^{\text{ME}%
}=0  \label{S1}
\end{equation}%
for $\mu >0$ up to the first order in $J$ and $\lambda $, which is shown as
the blue lines in Fig.\ref{FigEdel}. 

\begin{figure}[t]
\centerline{\includegraphics[width=0.48\textwidth]{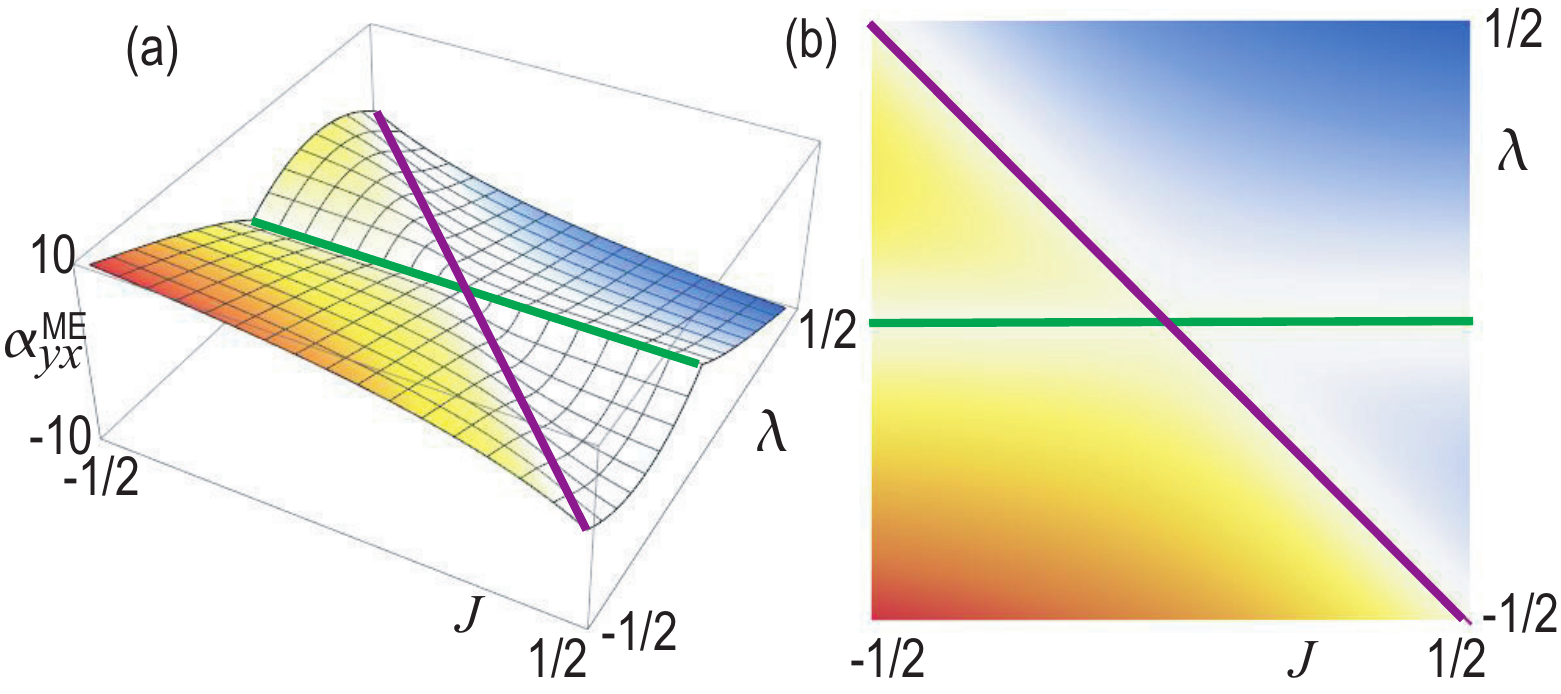}}
\caption{In-plane Edelstein effect. Bird's eye's view of $\protect\alpha %
_{yx}^{\text{ME}}$ in the $J$-$\protect\lambda $ plane, and (b) its color
plot. Green line represents the condition $\protect\lambda =0$ and the
purple line represents the condition $\protect\lambda +J=0$. See the caption
of Fig.2 for the units of various variables. We have set $\protect\mu =2%
\protect\varepsilon _{0}$ and $m=4\hbar ^{2}k_{0}^{2}/\protect\varepsilon %
_{0}$. }
\label{FigPhase}
\end{figure}

We show the numerically obtained susceptibility in Fig.\ref{FigEdel}. Eq.(%
\ref{S1}) agrees well with the numerical result for small $J/\lambda $. The
susceptibility is zero $\alpha _{ix}^{\text{ME}}=0$ for $\lambda =0$ or $%
J+\lambda =0$ as shown in Fig.\ref{FigEdel}(a) and (b), where the
Hamiltonian (\ref{pHamil})\ is diagonal. The susceptibility $\alpha _{yx}^{%
\text{ME}}$ is positive for $J+\lambda <0$ and negative for $J+\lambda >0$
as shown in in Fig.\ref{FigPhase}(a) and (b). It is because the spin
direction along the $x$ axis matters for the electric-field induced
magnetization. On the other hand, the sign of the susceptibility $\alpha
_{yx}^{\text{ME}}$ does not change at $\lambda =0$ because $f^{\left(
1\right) }\left( 0,k_{y}\right) =0$.

\textbf{Discussion:} We investigated the electric-field induced
magnetization in a simple two-band $p$-wave magnet model, where the Rashba
interaction is essential. On the other hand, the electric-field induced
magnetization was shown to occur without the Rashba interaction in the $p$%
-wave magnet system described by the four-band model\cite{pwave,Chak} 
\begin{eqnarray}
H &=&2t\left( \tau _{x}\cos \frac{k_{x}}{2}+\cos k_{y}\right)  \notag \\
&&+2t_{J}\left( \sigma _{x}\tau _{y}\sin \frac{k_{x}}{2}+\sigma _{y}\tau
_{z}\cos k_{y}\right) ,  \label{CNAO}
\end{eqnarray}%
where $\sigma $ and $\tau $ correspond to the spin and site degrees of
freedom.

The above four-band model is derived for a specific material candidate
CeNiAsO\cite{pwave}, which is a metal. The dependence of the pseudospin $%
\tau $\ enters in a specific way, and hence, this four-band Hamiltonian is
applicable to a specific case. On the other hand, the two-band Hamiltonian (%
\ref{pHamil}) is quite universal. There are two possibilities to realize the
model. One is the metallic $p$-wave magnet without the pseudospin degrees of
freedom on a substrate producing the Rashba interaction. The other is the
insulating $p$-wave magnet, where free electrons exist away from the $p$%
-wave magnet, while the interface produces the Rashba interaction. Our
results will open a new way to multiferroics based on the $p$-wave magnet.

This work is supported by CREST, JST (Grants No. JPMJCR20T2) and
Grants-in-Aid for Scientific Research from MEXT KAKENHI (Grant No. 23H00171).


\begin{thebibliography}{99}
\bibitem{Fiebig} M. Fiebig, Revival of the magnetoelectric effect, J. Phys.
D: Appl. Phys. 38 R123 (2005).

\bibitem{Eeren} W. Eerenstein, N. D. Mathur and J. F. Scott, Multiferroic
and magnetoelectric materials, Nature 442, 759 (2006).

\bibitem{Ramesh} R. Ramesh and Nicola A. Spaldin, Multiferroics: progress
and prospects in thin films Nature Materials 6, 21 (2007).

\bibitem{Kimura} T. Kimura, T. Goto, H. Shintani, K. Ishizaka, T. Arima and
Y. Tokura, Magnetic control of ferroelectric polarization, Nature 426, 55
(2003).

\bibitem{Hur} N. Hur, S. Park, P. A. Sharma, J. S. Ahn, S. Guha and S-W.
Cheong, Electric polarization reversal and memory in a multiferroic material
induced by magnetic fields, Nature 429, 392 (2004).

\bibitem{Nan} Ce-Wen Nan, Gang Liu, and Yuanhua Lin and Haydn Chen,
Magnetic-Field-Induced Electric Polarization in Multiferroic Nanostructures,
Phys. Rev. Lett. 94, 197203 (2005).

\bibitem{Ohno} H. Ohno, D. Chiba, F. Matsukura, T. Omiya, E. Abe, T. Dietl,
Y. Ohno and K. Ohtani, Electric-field control of ferromagnetism, Nature 408,
944 (2000).

\bibitem{Chu} Ying-Hao Chu, Lane W. Martin, Mikel B. Holcomb, Martin Gajek,
Shu-Jen Han, Qing He, Nina Balke, Chan-Ho Yang, Donkoun Lee, Wei Hu, Qian
Zhan, Pei-Ling Yang, Arantxa Fraile-Rodrguez, Andreas Scholl, Shan X. Wang
and R. Ramesh, Electric-field control of local ferromagnetism using a
magnetoelectric multiferroic, Nature Materials 7, 478 (2008).

\bibitem{Heron} J. T. Heron, M. Trassin, K. Ashraf, M. Gajek, Q. He, S. Y.
Yang, D. E. Nikonov, Y-H. Chu, S. Salahuddin and R. Ramesh,
Electric-Field-Induced Magnetization Reversal in a Ferromagnet-Multiferroic
Heterostructure, Phys. Rev. Lett. 107, 217202 (2011).

\bibitem{Matsukura} Fumihiro Matsukura, Yoshinori Tokura, Hideo Ohno,
Control of magnetism by electric fields, Nature Nanotechnology 10, 209 (2015).

\bibitem{Edel} V.M. Edelstein, Spin polarization of conduction electrons
induced by electric current in two-dimensional asymmetric electron systems,
Solid State Communications 73 (3): 233 (1990).

\bibitem{Gamba} P. Gambardella and I. M. Miron, Current-induced spin-orbit
torques, Philos. Trans. R. Soc. A Math. Phys. Eng. Sci. 369, 3175 (2011).

\bibitem{Manchon} A. Manchon, H. C. Koo, J. Nitta, S. M. Frolov, and R. A.
Duine, New perspectives for rashba spin-orbit coupling, Nat. Mater. 14, 871
(2015).

\bibitem{Sanchez} J. C. Rojas-Sanchez, L. Vila, G. Desfonds, S. Gambarelli,
J.P. Attane, J. M. De Teresa, C. Magen, A, Fert, Spin-to-charge conversion
using Rashba coupling at the interface between non-magnetic materials,
Nature Communications 4, 2944 (2013).

\bibitem{Kawasuso} H.\thinspace J. Zhang, S. Yamamoto, B. Gu, H. Li, M.
Maekawa, Y. Fukaya, and A. Kawasuso, Charge-to-Spin Conversion and Spin
Diffusion in Bi/Ag Bilayers Observed by Spin-Polarized Positron Beam, Phys.
Rev. Lett. 114, 166602 (2015).

\bibitem{Du} Ye Du, Hiromu Gamou, Saburo Takahashi, Shutaro Karube, Makoto
Kohda, and Junsaku Nitta, Disentanglement of Spin-Orbit Torques in Pt/Co
Bilayers with the Presence of Spin Hall Effect and Rashba-Edelstein Effect,
Phys. Rev. Applied 13, 054014 (2020).

\bibitem{Kato} Y. Kato, R. C. Myers, A. C. Gossard, and D. D. Awschalom,
Coherent spin manipulation without magnetic fields in strained
semiconductors, Nature 427, 50 (2004).

\bibitem{Peters} R. Peters and Y. Yanase, Strong enhancement of the
Edelstein effect in $f$-electron systems, Phys. Rev. B 97, 115128 (2018).

\bibitem{Bryk} V. V. Bryksin and P. Kleinert, Theory of
electric-field-induced spin accumulation and spin current in the
two-dimensional Rashba model, Phys. Rev. B - Condens.Matter Mater. Phys. 73,
1 (2006).

\bibitem{Monte} S. Leiva-Montecinos, J. Henk, I. Mertig, and A. Johansson,
Spin and orbital Edelstein effect in a bilayer system with Rashba
interaction, Phys. Rev. Res. 5, 1 (2023).

\bibitem{Mell} A. R. Mellnik, J. S. Lee, A. Richardella, J. L. Grab, P. J.
Mintun, M. H. Fischer, A. Vaezi, A. Manchon, E.-A. Kim, N. Samarth \& D. C.
Ralph Spin-transfer torque generated by a topological insulator, Nature
volume 511, 449 (2014).

\bibitem{SanchezPRL} J.-C. Rojas-Sanchez, S. Oyarzun, Y. Fu, A. Marty, C.
Vergnaud, S. Gambarelli, L. Vila, M. Jamet, Y. Ohtsubo, A. Taleb-Ibrahimi,
P. Le Fevre, F. Bertran, N. Reyren, J.-M. George, A. Fert, Spin-pumping into
surface states of topological insulator $\alpha $-Sn, spin to charge
conversion at room temperature, Phys. Rev. Lett. 116, 096602 (2016).

\bibitem{Johan} A. Johansson, J. Henk, and I. Mertig, Edelstein effect in
Weyl semimetals, Phys. Rev. B 97, 1 (2018).

\bibitem{pwave} Anna Birk Hellenes, Tomas Jungwirth, Jairo Sinova, Libor 
\v{S}mejkal, Unconventional p-wave magnets, arXiv:2309.01607.

\bibitem{Chak} Atasi Chakraborty, Anna Birk Hellenes, Rodrigo
Jaeschke-Ubiergo, Tomas Jungwirth, Libor \v{S}mejkal, Jairo Sinova, Highly
Efficient Non-relativistic Edelstein effect in p-wave magnets,
arXiv:2411.16378.

\bibitem{Garete} I. Garate, K. Gilmore, M. D. Stiles, and A. H. MacDonald,
Nonadiabatic spin-transfer torque in real materials, Phys. Rev. B 79, 104416
(2009).

\bibitem{Garete2} Ion Garate, A.H. MacDonald, Influence of a Transport
Current on Magnetic Anisotropy in Gyrotropic Ferromagnets, Phys. Rev. B 80,
134403 (2009).

\bibitem{Zele} J. Zelezny, H. Gao, K. Vyborny, J. Zemen, J. Masek, A.
Manchon, J. Wunderlich, J. Sinova, and T. Jungwirth, Relativistic Neel-order
fields induced by electrical current in antiferromagnets, Phys. Rev. Lett.
113, 1 (2014).

\bibitem{Li2015} Hang Li, H. Gao, Liviu P. Zarbo, K. Vyborny, Xuhui Wang,
Ion Garate, Fatih Dogan, A. Cejchan, Jairo Sinova, T. Jungwirth, Aurelien
Manchon, Intraband and interband spin-orbit torques in noncentrosymmetric
ferromagnets, Phys. Rev. B 91, 134402 (2015).

\bibitem{ManchonRMP} A. Manchon, J. Zelezny, I. M. Miron, T. Jungwirth, J.
Sinova, A. Thiaville, K. Garello, and P. Gambardella, Current-induced
spin-orbit torques in ferromagnetic and antiferromagnetic systems, Reviews
of Modern Physics 91, 035004 (2019).

\bibitem{Tenzin} Karma Tenzin, Arunesh Roy, Frank T. Cerasoli, Anooja
Jayaraj, Marco Buongiorno Nardelli, Jagoda S\l awi\'{n}ska, Collinear
Rashba-Edelstein effect in non-magnetic chiral materials, Phys. Rev. B 108,
245203 (2023).

\bibitem{Her} R. Gonzalez-Hernandez, P. Ritzinger, K. Vyborny, J. Zelezny,
and A. Manchon, Non-relativistic torque and edelstein effect in
non-collinear magnets, Nature Communications 15, 1 (2024).

\bibitem{Chen} Y. Chen, Z. Z. Du, H.-Z. Lu, and X. C. Xie,
Intrinsicspin-orbit torque mechanism for deterministic all-electric
switching of noncollinear antiferromagnets, Phys. Rev. B 109, L121115 (2024).

\bibitem{Trama} Mattia Trama, Irene Gaiardoni, Claudio Guarcello, Jorge I.
Facio, Alfonso Maiellaro, Francesco Romeo, Roberta Citro, Jeroen van den
Brink, Non-linear anomalous Edelstein response at altermagnetic interfaces,
arXiv:2410.18036.

\bibitem{Hu} Mengli Hu, Oleg Janson, Claudia Felser, Paul McClarty, Jeroen
van den Brink, Maia G. Vergniory, Spin Hall and Edelstein Effects in Novel
Chiral Noncollinear Altermagnets, arXiv:2410.17993.

\bibitem{Pari} Nayra A. Alvarez Pari, Rodrigo Jaeschke-Ubiergo, Atasi
Chakraborty, Libor Smejkal and Jairo Sinova, Non-relativistic linear
Edelstein effect in non-collinear EuIn$_{2}$ As$_{2}$, arXiv:2412.10984.

\bibitem{EzawaPNeel} M. Ezawa, Purely electrical detection of the Neel
vector of p-wave magnets based on linear and nonlinear conductivities,
arXiv:2410.21854.

\bibitem{SmejRev} L. Smejkal, A. H. MacDonald, J. Sinova, S. Nakatsuji and
T. Jungwirth, Anomalous Hall antiferromagnets, Nat. Rev. Mater. 7, 482
(2022).

\bibitem{SmejX2} Libor \v{S}mejkal, Jairo Sinova, and Tomas Jungwirth,
Emerging Research Landscape of Altermagnetism, Phys. Rev. X 12, 040501
(2022).

\bibitem{Zu2023} Di Zhu, Zheng-Yang Zhuang, Zhigang Wu, and Zhongbo Yan,
Topological superconductivity in two-dimensional altermagnetic metals, Phys.
Rev. B 108, 184505 (2023).

\bibitem{Sun} Chi Sun, Jacob Linder, Spin pumping from a ferromagnetic
insulator into an altermagnet, Phys. Rev. B 108, L140408 (2023).

\bibitem{Diniz} G. S. Diniz and E. Vernek, Suppressed Kondo screening in
two-dimensional altermagnets, Phys. Rev. B 109, 155127 (2024).

\bibitem{Rao} Peng Rao, Alexander Mook, Johannes Knolle, Tunable band
topology and optical conductivity in altermagnets, Phys. Rev. B 110, 024425
(2024).

\bibitem{Amund} Morten Amundsen, Arne Brataas, Jacob Linder, RKKY
interaction in Rashba altermagnets, Phys. Rev. B 110, 054427 (2024).

\bibitem{EzawaAlter} M. Ezawa. Detecting the Neel vector of altermagnets in
heterostructures with a topological insulator and a crystalline valley-edge
insulator, Phys. Rev. B 109 (24), 245306 (2024).

\bibitem{Maeda} Kazuki Maeda, Bo Lu, Keiji Yada, Yukio Tanaka, Theory of
tunneling spectroscopy in p-wave altermagnet-superconductor hybrid
structures, J. Phys. Soc. Jpn. 93, 114703 (2024).

\bibitem{EzawaPwave} M. Ezawa, Topological insulators based on $p$-wave
altermagnets; Electrical control and detection of the altermagnetic domain
wall, Phys. Rev. B 110, 165429 (2024).

\bibitem{Brek} Bjonulf Brekke, Pavlo Sukhachov, Hans Glokner Giil, Arne
Brataas, Jacob Linder, Minimal models and transport properties of
unconventional p-wave magnets, Phys. Rev. Lett. 133, 236703 (2024).

\bibitem{GIAlter} M. Ezawa, Third-order and fifth-order nonlinear
spin-current generation in g-wave and i-wave altermagnets and perfect
spin-current diode based on f-wave magnets, arXiv:2411.16036.
\end{thebibliography}
\end{document}